\documentstyle[preprint,aps]{revtex}
\begin{document}
\preprint{OITS-553}
\draft
\title{Isospin Structure of Penguins \\And Their Consequences in $B$ Physics
\footnote{Work supported in part by the Department of Energy Grant No.
DE-FG06-85ER40224.}}
\author{N.G. Deshpande and
Xiao-Gang He}
\address{Institute of Theoretical Science\\
University of Oregon\\
Eugene, OR 97403-5203, USA}
\date{August, 1994}
\maketitle
\begin{abstract}
Isospin structure of gluon mediated or strong penguin is significantly altered
when the full electroweak corrections are included. This has the consequence
that some previous analyses which relied on a simple isospin structure in
charmless $B$ decays become inapplicable. We present the general
Hamiltonian in next-to-leading order QCD, and illustrate our conclusion
quantitatively for $B\rightarrow \pi\pi$ and $B\rightarrow K\pi$ decays in the
factorization approximation. Some remarks on CP
asymmetries in $B$ decays are also made.

\end{abstract}
\pacs{12.15.Ji, 12.38.Bx, 13.20.He, 13.25.Hw}
\newpage

Flavor changing one loop processes mediated by a gluon play an important role
in understanding the Standard Model (SM)\cite{1,2,3,4}. The CP violating
parameter $\epsilon'/\epsilon$ in the K system is estimated by evaluating the
flavor changing process $s\rightarrow d q \bar q$, where $q = u$ or $d$. It is
by now well known that substantial corrections arise in this estimate from
inclusion of the full electroweak effects\cite{5,6} (the $\gamma$, $Z$ penguins
and the ``box'' contributions). These additional electroweak corrections add
$\Delta I = 3/2$
contribuitons to the $\Delta I = 1/2$ pure gluon exchange, and substantially
reduce the estimate for $\epsilon'/\epsilon$. A complete next-to-leading order
(NLO) calculation now exists for the full electroweak contributions for such
flavor changing process\cite{6}. In the K system the enhancement of $\Delta I =
1/2$ amplitude over $\Delta I =3/2$ by a factor of about 20 in $K\rightarrow
\pi\pi$ decays further amplifies the electroweak contribution to
$\epsilon'/\epsilon$.
One might think that in the $B$ system electroweak corrections to
gluon exchange may be negligible. This in fact is incorrect. The large mass of
the top increases the relative contributions of the electroweak corrections and
has the effect of causing significant reduction in the estimates for gluon
mediated decays. Our recent estimates for $b\rightarrow s\phi$ process suggests
a reduction in the rate of 20\%$\sim$ 30\%\cite{3,4}. The gluon mediated
process $b\rightarrow s q \bar q$ where $q = u\;, d\;, s$ is pure $\Delta I =
0$. Inclusion of the full electroweak corrections will introduce a significant
$\Delta I = 1$ admixture. Any analysis that relies on penguin contributions
being pure $\Delta I = 0$ is therefore suspect. We will comment on an
interesting analysis assuming the above property which was recently published
in Physical Review Letters\cite{7}. Similarly, one might worry about the wrong
isospin admixture in $b\rightarrow d q\bar q$ process where the gluon exchange
is pure $\Delta I = 1/2$. Extraction of penguin effects from the tree effects
are important in $B\rightarrow \pi\pi$ process\cite{8,9} because of the need to
remove penguin contamination to make a measurement of the angle $\alpha$ in the
unitarity
triangle of the KM matrix.

We shall first present the full Hamiltonian describing flavor changing
processes in the NLO approximation. The estimate of matrix elements to quantify
our results can only be done in the context of a model in our present
calculational limitation. We shall use factorization approximation and some
models for form factors when necessary.

The effective Hamiltonian responsible for charmless $B$ decays can be
parametrized as

\begin{eqnarray}
H_{\Delta B=1} = {G_F\over \sqrt{2}}[V_{ub}V^*_{uq}(c_1O^u_1 + c_2 O^u_2)
 - V_{tb}V^*_{tq}\sum^{10}_{i=3} c_iO_i] +H.C.\;,
\end{eqnarray}
where the Wilson coefficients (WCs) $c_i$ are defined at the scale of $\mu
\approx
m_b$, and $O_i$ are defined as
\begin{eqnarray}
O^u_1 = \bar q_\alpha \gamma_\mu(1-\gamma_5)u_\beta\bar
u_\beta\gamma^\mu(1-\gamma_5)b_\alpha\;&,&\;\;\;\;
O^u_2 = \bar q \gamma_\mu(1-\gamma_5)u\bar
u\gamma^\mu(1-\gamma_5)b\;,\nonumber\\
O_3 = \bar q \gamma_\mu(1-\gamma_5)b \sum_{q'}
\bar q' \gamma_\mu(1-\gamma_5) q'\;&,&\;\;\;\;
Q_4 = \bar q_\alpha \gamma_\mu(1-\gamma_5)b_\beta \sum_{q'}
\bar q'_\beta \gamma_\mu(1-\gamma_5) q'_\alpha\;,\nonumber\\
O_5 =\bar q \gamma_\mu(1-\gamma_5)b \sum_{q'} \bar q'
\gamma^\mu(1+\gamma_5)q'\;&,&\;\;\;\;
Q_6 = \bar q_\alpha \gamma_\mu(1-\gamma_5)b_\beta \sum_{q'}
\bar q'_\beta \gamma_\mu(1+\gamma_5) q'_\alpha\;,\\
O_7 ={3\over 2}\bar q \gamma_\mu(1-\gamma_5)b \sum_{q'} e_{q'}\bar q'
\gamma^\mu(1+\gamma_5)q'\;&,&\;\;
Q_8 = {3\over 2}\bar q_\alpha \gamma_\mu(1-\gamma_5)b_\beta \sum_{q'}
e_{q'}\bar q'_\beta \gamma_\mu(1+\gamma_5) q'_\alpha\;,\nonumber\\
O_9 ={3\over 2}\bar q \gamma_\mu(1-\gamma_5)b \sum_{q'} e_{q'}\bar q'
\gamma^\mu(1-\gamma_5)q'\;&,&\;\;
Q_{10} = {3\over 2}\bar q_\alpha \gamma_\mu(1-\gamma_5)b_\beta \sum_{q'}
e_{q'}\bar q_\beta \gamma_\mu(1-\gamma_5) q'_\alpha\;.\nonumber
\end{eqnarray}
Here $q$ can be $d$ or $s$, and $O_2$, $O_1$ are the tree level
and QCD corrected operators. $O_{3-6}$ are the strong
gluon induced operators, and operators $O_{7-10}$ are due to $\gamma$ and $Z$
exchange, and ``box'' diagrams at loop level. We refer to the latter as
electroweak penguin operators. The summation on $q'$ in all the operators are
over $u\;,\;d\;,\;s$ quarks. The $\sum_{q'}$ terms in $Q_{3-6}$ transform as
isosinglet, while the presence of $e_{q'}$ in $Q_{7-10}$ makes the
$\sum_{q'}e_{q'}$ terms transform as a mixture of $I = 0$ and $I=1$ components.
Consequently,
 $O_{1,2}$ have two isospin components, the strong penguin  has a single
isospin
component, and the electroweak penguin again contains two isospin components.

 Naively, the electroweak penguin contribuiton compared
with the strong penguin contribution is suppressed by a factor
$\alpha_{em}/\alpha_s$ and therefore one might be tempted to neglect them. For
large top quark mass, this is, however, no longer true
because there is a term in the electroweak penguin contribution in which the WC
is proportional to the squre of the top quark mass.

The WCs $c_i$ at the scale $\mu = m_b$ are obtained by first
calculating the WCs at $\mu = m_W$ and then
using
the renormalization group equation to evolve them down to $m_b$. We carry out
this analysis using the next-to-leading order QCD corrected WCs following
Ref\cite{6}.  We use in our analysis, $\alpha_s(m_Z)=0.118$, $\alpha_{em}(m_Z)
= 1/128$, $m_t = 174$ GeV and $m_b = 5$ GeV, and obtain\cite{3}
\begin{eqnarray}
c_1 &=& -0.3125\;, \;\;c_2 = 1.1502\;, \;\;c_3 = 0.0174\;, \;\;c_4 = -0.0373\;,
\nonumber\\
c_5 &=& 0.0104\;,
\;\;c_6 = -0.0459\;,\;\; c_7 = -1.050\times 10^{-5}\;,\nonumber\\
 c_8 &=& 3.839\times 10^{-4}\;,
\;\;c_9 = -0.0101\;, \;\;c_{10} = 1.959\times 10^{-3}\;.
\end{eqnarray}
We see that the coefficient $c_9$ arising from electroweak penguin contribution
is not much smaller
than coefficients of the strong penguin. The major contribution to $c_9$ arises
from $Z$ penguin. The dependence of $c_i$ on $m_t$ is given in Ref.\cite{3}.

Our coefficients are given in a regularization independent scheme which
requires that the matrix elements
be renormalized to one loop level for consistency.
These one-loop matrix elements can be rewritten in terms of the tree-level
matrix elements $<O_j>^{tree}$ of the effective operators, and one
obtains\cite{3,4,10}
\begin{eqnarray}
< c_i O_i> = \sum_{ij} c_i(\mu) [\delta_{ij} +{\alpha_s\over 4\pi}m^s_{ij}
+{\alpha_{em}\over 4\pi}m^e_{ij}] <O_j>^{tree} = \sum c'_i<Q_i>^{tree}\;.
\end{eqnarray}
Here $m^{s,e}$ are $10\times 10$ matrices which we have evaluated.
Expressing the effective coefficients $c'_i$ which multiply
the matrix elements
$<O_i>^{tree}$ in terms of $ c_i$, we have
\begin{eqnarray}
c'_1 &=& c_1\;\;\;\;\;\;\;\; c'_2 = c_2\nonumber\\
c'_3 &=&  c_3 - P_s/3\;,\;\; c'_4 =  c_4 +P_s\;,\;\;
c'_5 = c_5 - P_s/3\;,\;\;c'_6 = c_6 + P_s\;,\nonumber\\
c'_7 &=&  c_7 +P_e\;,\;\;\;\;\;\;c'_8 =
c_8\;,\;\;\;\;\;\;\;\;\;\;
c'_9 =   c_9 +P_e\;,\;\;\;\;\; c'_{10} =  c_{10}\;.
\end{eqnarray}
The leading contributions to $P_{s,e}$ are given by:
 $P_s = (\alpha_s/8\pi) c_2 (10/9 +G(m_c,\mu,q^2))$ and
$P_e = (\alpha_{em}/9\pi)(3 c_1+ c_2) (10/9 + G(m_c,\mu,q^2))$. Here
$m_c$ is the charm quark mass which we take to be 1.35 GeV. The function
$G(m,\mu,q^2)$ is give by
\begin{eqnarray}
G(m,\mu,q^2) = 4\int^1_0 x(1-x) \mbox{d}x \mbox{ln}{m^2-x(1-x)q^2\over
\mu^2}\;,
\end{eqnarray}
and $q^2$ is the momentum of the exchanged particle. In factorization
approximation, the final state phase only arises from the imaginary part of G.

\newpage
\noindent
{\bf a) $B\rightarrow \pi\pi$ and extraction of $\mbox{sin}2\alpha$}.

In this case $q=d$, and the effective Hamiltonian contains $\Delta I = 1/2$
and $\Delta I = 3/2$ amplitudes. The $\Delta I = 1/2$ component will contribute
to decay amplitudes with $I = 0$ and $ I = 1$ in the final states, while the
$\Delta I = 3/2$ component will contribute to the amplitudes with $I = 1$ and 2
in the final states. However, Bose symmetry requires $\pi\pi$ to be in $I = 0$
or $I=2$ state. The KM matrix elements in front of $O_{1,2}$, the strong
penguin, and the electroweak penguin contributions are comparable, while the
WCs of the strong penguin and the electroweak penguin are
all smaller than the ones for $O_{1,2}$.

The extraction of $\alpha$, one of three angles of the unitarity triangle
defined
by the KM matrix elements, involves the study of CP asymmetry in $B \rightarrow
\pi\pi$ and $\bar B\rightarrow \pi\pi$ channels. If the pegnuin contributions
are
neglected, this extraction is straightforward. However, if penguin diagrams
make a significant contribution, then the interpretations of the results become
complicated. An isospin analysis of $B\rightarrow \pi\pi$ has been presented by
Gronau and London\cite{8}, where a method of removing penguin contribution was
provided. An important assumption in this method is that the $\Delta I = 3/2$
amplitude involved in $B^\pm \rightarrow \pi^\pm \pi^0$ decays arises purely
from the operators $O_{1,2}$. Since we now have the electroweak penguin
operators with $\Delta I = 3/2$ part, one can ask if this method is still
valid. We calculate the ratio of the penguin to the tree amplitude in the
factorization approximation,
\begin{eqnarray}
A_{tree}(B^+\rightarrow \pi^+\pi^0)&=& -{G_F\over \sqrt{2}} V_{ub}^*V_{ud}
(c'_1+ c'_2)(1+\xi) T\;,\nonumber\\
A_{penguin}(B^+\rightarrow \pi^+\pi^0)&=&  {G_F\over \sqrt{2}} V_{tb}^*V_{td}
{3\over 2}[c'_7+\xi c'_8+c'_9+\xi c'_{10} + \xi c'_9+c'_{10}\nonumber\\
&+& {2\over3} (\xi c'_7+c'_8)(2W+X)] T\;,
\end{eqnarray}
where $\xi=1/N$ with N being the number of colors, $W =
m_\pi^2/(m_d+m_u)(m_b-m_d)$, $X = m_\pi^2/2m_d(m_b-m_d)$, and $T =
if_{\pi^0}(f^+_{B\pi^-}(m_B^2-m_\pi^2)
+f^-_{B\pi^-}m_\pi^2)$. Here $f_{\pi^0}= 93$ MeV is the pion form factor. For
our numerical calculations we will use $m_u = 5.7$ MeV, $m_d = 8.7$ MeV, and
the form factors $f^\pm$ calculated in
Ref.\cite{11,12}. We will treat $\xi$ as
a parameter and use the experimental data favored number $1/2$\cite{12}.

In factorization approximation, the ratio of the tree and penguin amplitudes do
not depend on the form factors. We find
\begin{eqnarray}
{|A_{penguin}(B^+ \rightarrow \pi^+\pi^0)|\over |A_{tree}(B^+ \rightarrow
\pi^+\pi^0)|} \approx 1.6\% (|V_{td}|/ |V_{ub}|).
\end{eqnarray}
Here we have neglected a small contribution about 0.5\% due to strong penguin
from isospin breaking effect for $m_d \neq m_u$.

We see that the assumption made in Ref.\cite{8} is quite good.
We have also calculated
the ratio of penguin to tree amplitudes for $B^0\rightarrow \pi^+\pi^-$ and
$B^0\rightarrow \pi^0\pi^0$, and find them to be about $7\%(|V_{td}|/|V_{ub}|)$
and
$23\%(|V_{td}|/|V_{ub}|)$, respectively. The larger penguin effect in
$B^0\rightarrow \pi^0\pi^0$ can be understood from the color suppression of the
tree contributions to this process in factorization approximation.

The method suggested in Ref.\cite{8} requires measurements of all the decay
amplitudes,
$B^+\rightarrow \pi^+\pi^0$, $B^0\rightarrow \pi^+\pi^-$, and
$B^0\rightarrow \pi^0\pi^0$. Experimentally it will be very difficult to
measure
the decay amplitude for $B^0\rightarrow \pi^0\pi^0$ to the desired accuracy for
a determination of $\alpha$. The CP asymmetry in $B^0\rightarrow \pi^+\pi^-$
decay
will be measured at the $B$ factory. We estimate below the error in $\alpha$
determination from this asymmetry in presence of penguin contamination. The
experimental measureable quatity is $\mbox{Im}\xi_{+-}$ where\cite{13}
\begin{eqnarray}
\xi_{+-} = e^{-2i\beta}{A(\bar B^0\rightarrow \pi^-\pi^+)
\over A(B^0\rightarrow \pi^+\pi^-)}\;,
\end{eqnarray}
and  $e^{-2i\beta} = V^*_{tb}V_{td}/V_{tb}V^*_{td}$.
The $B^0$ decay amplitude in general has the form, $e^{i\gamma}a_{tree} +
e^{-i\beta}e^{i\delta}a_{penguin}$. Here
$a_{tree}$ and $a_{penguin}$ are the tree and the penguin contributions,
respectively, $\gamma = Arg(V_{ub}^*V_{us})$, and $\delta$ is the difference
between the tree and penguin rescattering strong phases.
In the absence of penguin
contributions, $\mbox{Im}(\xi_{+-}) = -\mbox{sin}
2\alpha$. Including the penguin contributions, we find the deviation from
$-\mbox{sin}2\alpha$ to be
\begin{eqnarray}
\Delta (\mbox{sin}2\alpha) &\equiv& Im(\xi_{+-}) + \mbox{sin}2\alpha
\nonumber\\
&=& R{2\mbox{cos}\delta\mbox{sin}3\alpha -
2\mbox{cos}(\delta+\alpha)\mbox{sin}2\alpha -
R\mbox{sin}2\alpha-R\mbox{sin}4\alpha \over
1+R^2-2R\mbox{cos}(\delta+\alpha)}\;,
\end{eqnarray}
where $R = a_{penguin}/a_{tree}$. For $B^0 \rightarrow \pi^+\pi^-$ decay,
$R$ is about 7\%. We find that $|\Delta(\mbox{sin}2\alpha)| < 0.14$ for all
values of $\alpha$ and $\delta$. The error in $\alpha$ determination could be
quite large. For example for $\alpha = 45^0$ this error leads to a
determination of $\alpha$ off by $13^0$.\\
\\
\noindent{\bf b) Isospin analysis in $B\rightarrow K\pi$ decays.}

We now consider the general Hamiltonian with $q=s$. The operators $O_{1,2}$
transform as $\Delta I = 0$ and 1, the strong penguin operators $O_{3-6}$
transform as $\Delta I = 0$ and the electroweak penguin transform as $\Delta I
= 0$ and 1. The operators $O_1$ and $O_2$ are suppressed by a factors
$|V_{ub}^*V_{us}|/|V_{tb}^*V_{ts}| \approx 1/50$, and penguin operators
dominate in spite of smaller WCs. It was show recently that isospin analysis of
$B\rightarrow K\pi$ can determine the angle $\gamma$ in the
KM unitarity triangle\cite{7}. The idea proposed in Ref.\cite{7} was to isolate
$\Delta I = 1$ transition in $B\rightarrow
K\pi$ decays, and if one could assume that this amplitude arose from the
operator $O_{1,2}$, one would know its weak phase. The presence of electroweak
penguins, whose contribution to this process is comparable to $O_{1,2}$, makes
this idea entirely unworkable. We shall set up the general analysis of the
problem with the inclusion of electroweak penguin, and arrive at quantitative
estimates based on factorization approximation.

The decay amplitudes for the charged $B$ decay into a kaon and a pion have the
following isospin decomposition
\begin{eqnarray}
A(B^+\rightarrow K^0\pi^+) = \sqrt{{1\over 3}} A_{3/2} - \sqrt{{2\over
3}}A_{1/2}\;,\nonumber\\
A(B^+\rightarrow K^+\pi^0) = \sqrt{{2\over 3}} A_{3/2} + \sqrt{{1\over
3}}A_{1/2}\;,
\end{eqnarray}
where $A_i$ indicates the amplitude with $I = i$ in the final state.

Consider the combination with pure $I = 3/2$ final state,
\begin{eqnarray}
A(B^+\rightarrow K^0\pi^+) + \sqrt{2} A(B^+\rightarrow K^+\pi^0) = \sqrt{3}
A_{3/2}\;.
\end{eqnarray}
In general we expect $A_{i}$ to have the structure in Wolfenstein
parametrization,
\begin{eqnarray}
A_{i} = {G_F\over \sqrt{2}}|V_{ub}^*V_{us}|[\tilde a_{tree,i}
e^{i\gamma}e^{i\tilde \delta_{T,i}}
+\tilde a_{penguin,i} e^{i\tilde \delta_{P,i}}]\;,
\end{eqnarray}
where $\tilde \delta_{T(P),i}$ is the tree (penguin) strong rescattering phase.
We resort to the factorization approximation to estimate the magnitudes of the
tree and penguin contributions. We find
\begin{eqnarray}
\sqrt{3}A_{1/2} &=& {G_F\over \sqrt{2}}\{ V_{ub}^*V_{us}[-(c'_1+\xi c'_2)C'
- (\xi c'_1 +c'_2)(T'+3A')]\nonumber\\
&-&V_{tb}^*V_{ts}[-3(\xi c'_3+c'_4)(T'+A')
- 6(\xi c'_5+c'_6)
(YT' - ZA')\nonumber\\
&+&{3\over 2}(c'_7+\xi c'_8)C' + 6(\xi c'_7+c'_8)ZA'
-3(\xi c'_9 +c'_{10})A' - {3\over 2}(c'_9+\xi c'_{10})T'] \}\;,\nonumber\\
\sqrt{{3\over2}} A_{3/2} &=& {G_F\over \sqrt{2}} \{V_{ub}^*V_{us} [-(c'_1+\xi
c'_2)C' - (\xi  c'_1+c'_2)T']
-V_{tb}^*V_{ts}[{3\over 2}(c'_7+\xi c'_8)C' \nonumber\\
&-& 3(\xi c'_7+c'_8)YT'
-{3\over 2}(\xi c'_9+c'_{10})T' -{3\over 2}(c'_9+\xi c'_{10})C']\}\;,
\end{eqnarray}
where
\begin{eqnarray}
Y &=& {m_K^2\over (m_b-m_u)(m_s+m_u)}\;,\;\;\;\;
Z = {m_B^2\over (m_b+m_u)(m_s-m_u)}\;,\nonumber\\
T'&=& if_K(f^+_{B\pi^0} (m_B^2-m_\pi^2) +f^-_{B\pi^0}m_K^2)\;,\nonumber\\
C'&=& if_{\pi^0}(f^+_{BK} (m_B^2-m_K^2) +f^-_{BK}m_\pi^2)\;,\nonumber\\
A' &=& -if_B(f_{K\pi^0}^+(m_K^2-m_\pi^2)-f_{K\pi^0}^-m_B^2)\;.
\end{eqnarray}
Note that if penguin contributions are neglected, we would find in $SU(3)$
limit,
\begin{eqnarray}
\sqrt{{3\over 2}}A_{3/2} = {V_{us}\over V_{ud}} A(B^+\rightarrow \pi^+\pi^0)\;,
\end{eqnarray}
as assumed by Ref.\cite{7}.  However, we now find
\begin{eqnarray}
A_{1/2} &=& {G_F\over \sqrt{2}}|V_{ub}^*V_{us}| (-0.747 e^{i\gamma}e^{i\tilde
\delta_{T,1/2}} +
7.305 e^{i\tilde \delta_{P,1/2}}) (\mbox{GeV}^3)\;;\nonumber\\
A_{3/2} &=& {G_F\over \sqrt{2}}|V_{ub}^*V_{us}| (-1.055
e^{i\gamma}e^{i\tilde\delta_{T,3/2}} +
0.843 e^{i\tilde \delta_{P,3/2}}) (\mbox{GeV}^3)\;.
\end{eqnarray}
In the above we have used: $V_{us} = 0.221$, $|V_{ts}|=  |V_{cb}|$,
$|V_{ub}/V_{cb}| = 0.08$, $f_K = 158$ MeV, and $m_s = 0.175$ GeV, $f_B = 200$
MeV, and the form factors calculated in Ref.\cite{11}.  The first term is the
contribution from the operators $O_{1,2}$, and the second term is due to the
electroweak penguin. We have also carried out a calculation using the from
factors in Ref.\cite{12}. We find that the ratios of the tree and penguin
amplitudes are about the same as the previous ones although the magnitudes are
about 30\% larger in the latter case.
It is clear that the electroweak penguin contribution to $A_{3/2}$ is
comparable to the contribution from $O_{1,2}$. This makes the analysis of
Ref.\cite{7} invalid.

We would like to remark that $B^+\rightarrow K^+\pi^0$ may still be a good
place to look for CP violation\cite{10}. The CP asymmetry defined by the
following equation,
\begin{eqnarray}
A_{asy} \equiv {\Gamma (B^+ \rightarrow K\pi)
- \Gamma (B^- \rightarrow \bar K \bar \pi) \over \Gamma (B^+ \rightarrow K\pi)
+ \Gamma (B^- \rightarrow \bar K \bar\pi)}\;,
\end{eqnarray}
can be estimated from eq.(14). The quark level strong rescattering phases
 $\tilde \delta_{T,i}$ is zero, $\tilde \delta_{P,1/2} $ and $\tilde
\delta_{P,3/2}$ are estimated to be $13^0$ and $-0.23^0$ respectively, using
$q^2 = m_b^2/2$ in eq.(6) with either from factors in Ref.{11} or
Ref.\cite{12}.
We find the asymmetry $A_{asy}$
can be as large as 10\% depending on the angle $\gamma$. This
estimate relies on the quark level strong rescattering phases. There may be
large long distance contributions to the strong rescattering phases from the
final state hadron interaction. These contributions
may dilute the CP asymmetry estimated here\cite{14,15}. Unfortunately one does
 not know how to calculate these long distance contributions.

\end{document}